\documentclass[aps,prl,reprint,twocolumn,superscriptaddress,nofootinbib,longbibliography]{revtex4-2}

\usepackage{braket}
\usepackage{appendix}

\usepackage{amsmath,amssymb,graphicx,psfrag}
\usepackage[utf8]{inputenc}
\usepackage[colorlinks,linkcolor={blue},citecolor={blue},urlcolor={blue}]{hyperref}

\usepackage{amssymb}
\usepackage{bbm}
\usepackage{caption}

\usepackage{graphicx}
\usepackage{subcaption}
\usepackage[font=small,labelfont=bf,
   justification=justified,
   format=plain]{caption} 
   
\usepackage{times}
\usepackage[margin=0.75in]{geometry}

\begin{document}


\title{Fractional Contribution of Dynamical
and Geometric Phases in Quantum Evolution }


\author{Arun Kumar Pati}
\email{ak.pati@kiit.ac.in}
\affiliation{Centre for Quantum Technology, KIIT University, Bhubaneswar, Odisha, India}

\author{Vlatko Vedral}
\affiliation{Clarendon Laboratory, University of Oxford, Parks Road, Oxford OX1 3PU, United Kingdom}

\author{Erik Sj{\"o}qvist}
\affiliation{Department of Physics and Astronomy, Uppsala University, Box 516, SE-751 20 Uppsala, Sweden}


\date{\today}

\begin{abstract}
The fundamental division of the total quantum evolution phase into geometric and dynamical components is a central problem in quantum physics. Here, we prove a remarkably simple and universal law demonstrating that this partitioning is governed, at every instant, solely by a single geometric quantity: the Bargmann angle (Bures angle). This result provides a universally applicable and rigorous way to define the exact fraction of the total phase that is geometric versus dynamical in origin, thereby establishing a new quantitative link between the dynamics of quantum evolution and the geometry of the state space. This finding has immediate practical consequences, furnishing a real-time measure of the “geometricity” of an evolution for designing high-fidelity geometric quantum gates with optimized robustness, and opening new avenues for quantum speed limit and coherent control.
\end{abstract}

\maketitle
\section{Introduction}

The geometric phase is a cornerstone of modern quantum theory, revealing that the evolution of a quantum system is profoundly influenced by the geometry of the path traced in its state space, independent of the detailed dynamics \cite{berry, aha}. 
It manifests in optics as the Pancharatnam phase in polarization transformations \cite{pancha}.
Subsequently, the notion of geometric phase was generalized for non-cyclic evolutions \cite{bhanda} and a kinematic approach was proposed \cite{muku}. The geometric phase was also generalized for non-adiabatic, non-cyclic and general quantum evolutions \cite{akp,akp1}. The concept of geometric phase for mixed states in interferometry has been proposed \cite{erik}.

Since its inception, this concept has permeated diverse fields of physics, providing a unifying framework for understanding a vast range of phenomena.  Its influence is seen in condensed matter physics, where the Berry curvature is central to the theory of the quantum Hall effect \cite{niu} and the classification of topological materials. 
The discovery of the geometric phase, in particular, has reshaped our understanding of quantum theory and spurred the development of novel concepts in geometric quantum computation \cite{ekert}.
Most critically, it provides the foundation for holonomic quantum computation \cite{lidar}, where the intrinsic robustness of the geometric phase to control noise is harnessed to design fault-tolerant quantum gates, representing a leading strategy in the quest to build a scalable quantum computer \cite{guo}. 

 It is well-established that for any general quantum evolution, the total phase is the sum of two distinct components: a dynamical phase, dependent on the system's energy and the passage of time, and a geometric phase, which depends solely on the geometric path traced on the projective Hilbert space \cite{jsa}.  While the additive nature of these phases is fundamental, a deeper question has remained unanswered: what universal principle governs the interplay between these two components during the quantum evolution? As a system evolves, how is the total infinitesimal phase change,  partitioned between its dynamical and geometric parts? Lacking a clear answer, the relationship between the geometry of the evolutionary path and the character of the resulting phase has been understood primarily through global properties of the trajectory.

In this Letter, we uncover a remarkably simple and universal law that provides a precise, local answer to this question. We prove that the partitioning of the total phase is governed at every instant solely by a single geometric quantity: the Bargmann angle \cite{barg} (Bures angle) \cite{bures}. 
Our result shows a universally applicable and rigorous way to define the fraction of the total phase that is geometric versus dynamical in origin.  It bridges the gap between the geometry of quantum space and the dynamics of evolution in a new, quantitative way. This is precisely the kind of insight that opens up new avenues for both fundamental theory and practical technology.
Our result provides a real-time measure of the ``geometricity'' of an evolution, quantifying its robustness against dynamical fluctuations. Second, it furnishes a powerful new tool for quantum control. By monitoring the distance from the origin, one can now actively steer an evolution to not only reach a target state but to do so with a desired phase character. This may open up new pathways for designing high-fidelity geometric quantum gates with optimized robustness, enhancing protocols in quantum metrology, and deepening our fundamental understanding of the intimate connection between the geometry of quantum space and its dynamics. It also will add to our geometrical understanding of quantum computation itself \cite{Gu}. 

Thus, quantifying the fractional contribution of dynamical and geometric phases to the total phase represents more than a minor addition to quantum mechanics; it provides a powerful new lens through which to understand and engineer quantum systems.
We mention a few immediate and impactful applications. 
(i) Geometric quantum computation: This is a major field that aims to build fault-tolerant quantum gates using the geometric phase, which is naturally resilient to certain types of control errors.
Our result would provide a precise metric for the ``geometricity'' of a quantum gate. For example, if a CNOT gate has a geometric phase fraction of $0.98$, then that would be a direct, quantitative statement about its robustness.
(ii) Optimization: It would become a key parameter in optimization protocols. A control algorithm could be designed not just to perform a gate, but to perform it while maximizing the geometric phase fraction, thereby actively designing for robustness.
(iii) Quantum Metrology and Sensing:
Precision measurements often rely on detecting minute phase shifts (e.g., in atomic clocks or interferometers). The total phase is often noisy because the dynamical part is sensitive to environmental fluctuations.
One could design sensors that operate in a regime where the geometric phase fraction is maximized, leading to more stable and reliable measurements. (iv) Quantum Control:
Our findings would provide physicists with a new ``knob to turn'' when designing how to steer a quantum state from one point to another in the quantum state space. The goal might be to achieve a target state in the shortest time, thus providing new Quantum Speed Limit  while also ensuring a high geometric phase fraction for stability. 

\section{The Main Result}
Before we state our main results, we need to understand the geometry of quantum evolution, which can be described using the notion of the projective Hilbert space.
 The projective Hilbert space  $\cal P= \mathbb{P}(\mathcal{H})$ for an $N$-dimensional quantum system with a Hilbert space $\mathcal{H}$ is the set of all rays in $\mathcal{H}$. A ray is an equivalence class of non-zero vectors in $\mathcal{H}$, where two vectors $|\psi\rangle$ and $|\phi\rangle$ are considered equivalent if they differ only by a non-zero complex scalar multiple, i.e., 
$\text{Ray } [|\psi\rangle] = \{ c |\psi\rangle : c \in \mathbb{C}, c \neq 0 \} $.
The projective Hilbert space $\mathcal{P}$ is the quotient space
$\mathbb{P}(\mathcal{H}) = \frac{\mathcal{H} \setminus \{0\}}{\sim}$, where the equivalence relation $\sim$ is defined via
$|\psi\rangle \sim |\phi\rangle \quad \text{if and only if} \quad |\phi\rangle = c |\psi\rangle \text{ for some } c \in \mathbb{C}, c \neq 0 $.

In quantum mechanics, a system's physical state is not defined by a single vector $|\psi\rangle \in \mathcal{H}$, but by a ray $[|\psi\rangle]$.  The projective Hilbert space $\mathbb{P}(\mathcal{H})$ naturally incorporates this fundamental principle by identifying all state vectors that differ only by a non-zero complex factor. For an $N$-dimensional quantum system, the projective Hilbert space $\mathbb{P}(\mathcal{H})$ is isomorphic to the complex projective space $\mathbb{C}\mathbb{P}^{N-1}$.
 For a qubit, $\mathbb{P}(\mathcal{H})$ is the Bloch sphere, $\mathbb{C}\mathbb{P}^1 \cong S^2$.

Consider a quantum system with initial state $|\Psi(0)\rangle \in \cal H^N$ that evolves to $|\Psi(t)\rangle \in \cal H^N$. During the time evolution, the quantum system acquires a total phase $\Phi_T$ which is also known as the Pancharatnam phase. The total phase
can be written as sum of the dynamical phase  $\Phi_D$ and the geometric phase  $\Phi_G$. Here, we will show that the fractional contribution of the dynamical and the geometric phases to the total phase are governed by the Bargmann angle (geodesic distance) between initial and current states at time $t$. Let us consider the unitary time evolution of the initial state $|\Psi(0) \rangle$ as given by
\begin{equation}
   |\Psi(t) \rangle = U(t) |\Psi(0) \rangle = e^{-\frac{i}{\hbar} H t }  |\Psi(0) \rangle,
\end{equation}
where $H$ is the time-independent Hamiltonian of the system. Before we prove our main result, we need the following results.

 \noindent
{\bf Lemma 1:} The action of an unitary operator $U$ on any state admits the orthogonal decomposition as given by
\begin{align}
 U |\Psi  \rangle = \langle U \rangle |\Psi  \rangle  + \Delta U |{\bar \Psi}(U) \rangle,
 \end{align}
 where $|{\bar \Psi}(U) \rangle$ is orthogonal to $| \Psi  \rangle$, $\langle U  \rangle = \langle \Psi |U | \Psi \rangle $ and 
 $\Delta U^2$ is the uncertainty in the unitary operator in the state $|\Psi \rangle$  \cite{levy,pati,bag} defined as 
\begin{align}
\Delta U^2 &= 
 \langle U^{\dagger} U  \rangle - 
 \langle  U^{\dagger} \rangle \langle U \rangle  \\ \nonumber
 &= (1- |\langle \Psi | U| \Psi \rangle|^2 ).
 \end{align}
 For proof, see Appendix. 
 
 {\bf Theorem} (Unitary Evolution Decomposition):
Let $U(t)$ be a unitary operator and $|\Psi(0)\rangle \in {\cal H}^N$ an initial quantum state. The time evolved state at any later time can be decomposed as 
\begin{eqnarray}
|\Psi(t)\rangle = \cos \frac{S_0(t)}{2} e^{i \Phi} | \Psi(0) \rangle + \sin \frac{S_0(t)}{2}
| {\bar \Psi}_0(t) \rangle,
\label{eq:unidecomp}
\end{eqnarray}
where $S_0(t)$ is the Bargmann angle $|\langle\Psi(0)|\Psi(t)\rangle| = \cos \frac{S_0(t)}{2}$ and $\Phi = \arg \langle\Psi(0)|\Psi(t)\rangle$ is the Pancharatnam phase or the total phase. 

\noindent
{\bf Proof:} Let the unitary operator $U= U(t)$ and $| \Psi \rangle = | \Psi(0) \rangle$.  Using the Lemma and denoting the orthogonal component $|{\bar \Psi}(U) \rangle = |{\bar \Psi}(U(t)) \rangle = |{\bar \Psi}_0(t) \rangle$, we have 
$ |\Psi(t)  \rangle = \langle U(t) \rangle |\Psi(0)  \rangle  + \Delta U(t) |{\bar \Psi}_0(t) \rangle$. The average of the unitary $U(t)$ is
$\langle \Psi(0) |U(t) | \Psi(0) \rangle = |\langle \Psi(0)|\Psi(t)\rangle| e^{i \Phi} = \cos \frac{S_0(t)}{2} e^{i \Phi}$ and the proof follows.
Note that $|{\bar \Psi}(U) \rangle$ depends on the unitary via $U(t) |\Psi(0) \rangle$, which evolves in time (unless the evolution is trivial, e.g., $U(t)$ is a phase times identity).

The Bargmann angle $S_0(t)$ is  also the shortest path connecting the initial state and the state at time $t$.
Given its dynamic nature and fundamental role in quantifying quantum evolution, we call $S_0(t)$ as the time-dependent 
Bargmann angle. This also tracks the evolving geometric relationship between quantum states along a dynamical path, while distinguishing it from the static Bargmann angle between fixed states.
This shows that even though the quantum system  lives in a $N$-dimensional Hilbert space, the time-evolution dynamics can be 
described within a two-dimensional subspace spanned by $\{ | \Psi(0) \rangle,  | {\bar \Psi}_0(t) \rangle  \}$. This observation is of independent interest and can be exploited separately.

Currently, we understand that the total phase acquired by a quantum state is the sum of two distinct components:
\begin{equation}
\Phi_T = \Phi= \Phi_{D} + \Phi_{G},
\label{eq:decomp}
\end{equation}
where $\Phi_{D}$ is the dynamical phase  and $\Phi_{G}$ is the geometric phase.
The dynamical phase is given by 
\begin{align}
\Phi_D = - \frac{1}{\hbar} \int \langle \Psi(t) |H | \Psi(t) \rangle ~dt.
\end{align} 
This depends on the energy of the system and the duration of evolution which is a measure of ``how long'' the system ran.
The geometric Phase $\Phi_G $ is given by the integral over the connection-form 
\begin{align}
\Phi_G = i \int \langle \chi(t) | {\dot \chi(t)} \rangle ~dt,
\end{align} 
where $|\chi(t) \rangle$ is the ``reference-state'' and defined as \cite{akp,akp1}
$$ |\chi(t) \rangle = \frac{\langle \Psi(t) | \Psi(0) \rangle }{|\langle \Psi(t) | \Psi(0) \rangle | }  | \Psi(t) \rangle.$$
The geometric phase depends only on the geometry of the path traced by the state on the projective Hilbert space. The 
``reference-state'' for the unitary time-evolution can be expressed as 
\begin{align}
|\chi(t)\rangle = \cos \frac{S_0(t)}{2} | \Psi(0) \rangle + \sin \frac{S_0(t)}{2} e^{-i \Phi}
 | {\bar \Psi}_0(t) \rangle .
 \end{align}

Now, we show how the infinitesimal change in the dynamical phase and the geometric phase are connected to the weighted sum and difference of infinitesimal change in the total phase and orthogonal phase, respectively, during the quantum evolution. A quantitative measure of their relative contributions would provide a universal figure of merit to classify any quantum process. Using above definitions and Eq.~\eqref{eq:decomp},  our central result can be expressed as

\begin{align}
d\Phi_D  = \cos^2 \frac{S_0(t)}{2}  d \Phi(t) 
+ \sin^2 \frac{S_0(t)}{2}  d {\bar \Phi}, \\ \nonumber
d\Phi_G  = \sin^2 \frac{S_0(t)}{2}  d \Phi(t) 
- \sin^2 \frac{S_0(t)}{2}  d {\bar \Phi}.
\end{align} 
Here, $d {\bar \Phi}$ is the infinitesimal phase coming from the evolution of the orthogonal component and  
${\bar \Phi} = -i \int \langle {\bar \Psi_0(t)} | {\dot {\bar \Psi}_0(t)} \rangle ~dt $.
This shows that the fractional contributions of dynamical and geometric phases are completely governed by the instantaneous fidelity (Bargmann angle) and geometric separation between the initial quantum state and the state at time 
$t$. Our result also provides new insights to the instantaneous geometric ``progress'' of the quantum evolution relative to its starting point, making it particularly useful for studying quantum speed limits, optimal control, and the geometric structure of quantum dynamics. 

Interpretation of the Dynamical Phase Relation:  This relation shows that the total dynamical phase is a weighted average of two contributions: (i) $\cos^2(S_0/2)  d\Phi$: Contribution from the total phase weighted by overlap with initial state
 $\cos^2(S_0/2) = |\langle \Psi(0)|\Psi(t)\rangle|^2$ which is the Bargmann angle. When state is near initial state ($S_0 \approx 0$), this dominates.
(ii) $\sin^2(S_0/2)  d{\bar \Phi}$: Contribution from orthogonal component's phase evolution weighted by 
 $\sin^2(S_0/2) = 1 - |\langle \Psi(0)|\Psi(t)\rangle|^2$ which is infidelity. When state is far from initial state ($S_0 \approx \pi$), this dominates.

Physical Interpretation of the Geometric Phase Relation:  The relation 
$d\Phi_G = \sin^2(S_0(t)/2) (d\Phi - d{\bar \Phi})$
reveals that the infinitesimal geometric phase arises from a fundamental phase mismatch between the global evolution of the quantum state and the internal evolution of its orthogonal component, scaled by how far the system has explored Hilbert space from its initial configuration. Here, $d\Phi$ represents the infinitesimal Pancharatnam phase of the state, while $d{\bar \Phi}$ captures the infinitesimal phase accumulated by the orthogonal part $|{\bar \Psi}_0(t)\rangle$. The prefactor $\sin^2(S_0(t)/2)$, which equals the infidelity $1 - |\langle \Psi(0)|\Psi(t)\rangle|^2$, acts as a geometric weight ensuring that the geometric phase only becomes significant when the state has substantially evolved away from its origin. This elegantly expresses that true geometric phase is not merely a global phase change, but a gauge-invariant measure of the anholonomy due to the curvature of the quantum state space, emerging from the differential rotation between the state's components in a moving frame.
 When $S_0$ is small (i.e., the state is close to the initial state), $\sin^2 S_0/2 \approx 0$. Thus, little geometric phase accumulates.
This behavior underscores the idea that the geometry of quantum state space becomes more influential as the state evolves away from its starting point. The relation effectively measures how ``efficiently'' the phase mismatch between the total phase changes and orthogonal phase is transformed into geometric phases, with the efficiency determined by the instantaneous state separation.

\section{Geometric Phase Fraction}

The introduction of a geometric phase fraction (GPF) can provide a new tool for foundational quantum mechanics, moving our understanding beyond the simple additive nature of phase. By defining a universal, dimensionless metric, we can now classify the intrinsic character of any quantum evolution, regardless of the physical system or the timescale over which it occurs. This allows for a direct and meaningful comparison between vastly different processes, from the slow, adiabatic transport of an electron spin to the rapid, laser-driven gate on a superconducting qubit. The GPF quantifies, at every instant, the degree to which the evolution is governed by the robust geometry of its path versus the noise-susceptible dynamics of its Hamiltonian. It transforms our perspective from merely acknowledging two phase components to understanding their continuous, dynamic interplay as a fundamental property of the evolution itself. It reveals that the very nature of quantum evolution (whether it is more ``geometric'' or more ``dynamical'') is determined by simple geometric distance in Hilbert space.

The geometric phase fraction is defined as the ratio of the geometric phase accumulation rate to the phase miss match rate during quantum evolution:

\begin{align}
f_g(t) = \frac{d\Phi_G}{(d\Phi - d{\bar \Phi})} = \sin^2\left(\frac{S_0(t)}{2}\right),
\end{align}
where $S_0(t)$ is the instantaneous Bures angle between the initial state $|\Psi(0)\rangle$ and the state at time $t$.
It has following properties: (i) $f_g(t) \in [0, 1]$, (ii) Monotonicity, i.e.,  generally increases with $S_0(t)$ and 
(iii) This is also gauge invariant (both numerator and denominator shift equally under gauge transformations).
The geometric phase fraction quantifies how ``geometrically efficient'' a quantum evolution is at a given moment. For example, if 
$f_g(t) \approx 0$, then the evolution is dominated by dynamical phase (state near initial position).
If $f_g(t) \approx 1$, the evolution is dominated by geometric phase (state far from initial position).
For other intermediate values, we have mixed contribution from both geometric and dynamical phases.
Therefore, it provides a quantitative measure of how quantum evolution explores the curved geometry of state space, with
implications for quantum information processing, fundamental physics, and quantum control theory.
The concept highlights that not all quantum evolutions are equally ``geometric,'' and this geometric character has tangible consequences for quantum technologies and our understanding of quantum dynamics.

\section{Quantum Metric and Quantum Circuitousness}

 We know that when a quantum system evolves in time, it traces a path on the projective Hilbert space. The total distance travelled by the quantum state can be measured using the Fubini-Study distance. The Fubini-Study metric provides a natural geometric measure of quantum evolution by quantifying the instantaneous separation between quantum states in the projective Hilbert space. For a quantum system evolving from 
time $|\Psi(t)\rangle$ to $|\Psi(t+dt)\rangle$, we can define the infinitesimal Fubini-Study distance \cite{aa,ap} as given by
\begin{align}
dS^2 = 4 (1 - |\langle \Psi(t)|\Psi(t+dt) \rangle|^2 ) = 4 \Delta H^2 dt^2/\hbar^2,
\end{align}
where $\Delta H^2$ is the fluctuation in the Hamiltonian in the state $|\Psi(t)\rangle$. The total distance travelled by the state during an interval $[0, T]$ is given by
\begin{align}
S = 2 \int_0^T\Delta H dt/\hbar.
\end{align}
 This geometric perspective reveals that quantum dynamics can be viewed as a trajectory through curved state space.
 There is always a shortest possible path (geodesic) distance between the initial and the final states, 
while the actual evolution typically follows a longer path with accumulated dynamical and geometric phases. The difference between the total path length and the geodesic distance encodes information about the quantum system's dynamics, connecting to fundamental limits through the quantum speed limit and providing insights into optimal control strategies that minimize unnecessary evolution in the state space.

To make connection to the Fubini-Study metric and quantum speed limit, we go to a controlled-rotating frame where we can transform the
time-dependent orthogonal component to time-independent. In the orthogonal subspace, let $V(t)$ is a unitary that connects $| {\bar \Psi}_0(0) \rangle $ to $| {\bar \Psi}_0(t) \rangle$, i.e., $| {\bar \Psi}_0(t) \rangle = V(t) | {\bar \Psi}_0(0) \rangle$. In this frame, we can write the state at time $t$ as  
\begin{eqnarray}
|\Psi(t)\rangle = \cos \frac{S_0(t)}{2} e^{i \Phi} | \Psi(0) \rangle + \sin \frac{S_0(t)}{2}
 | {\bar \Psi}_0(0) \rangle.
 \end{eqnarray}
 
Now, using (13), we can find the infinitesimal Fubini-Study metric which is given by
\begin{eqnarray}
dS^2(t)  = dS_0(t)^2 + \sin S_0(t)^2 d\Phi(t)^2.
\label{eq:metric}
\end{eqnarray}
This is similar to the infinitesimal distance on the surface of a sphere (Bloch sphere for a qubit). 
The quantity $\sin^2 S_0(t)  d\Phi(t)^2$ represents the excess path length due to phase accumulation or the ``vertical motion'' in the $U(1)$ bundle. The geometric factor $\sin^2 S_0(t)$ reveals how far the system is from the initial state.
When $S_0 = 0$ (at initial state), we have $\sin^2 S_0 = 0$, thus no phase contribution to path length. 
When $S_0 = \pi/2$, we have $\sin^2 S_0 = 1 $, then maximum phase contribution appears.
Thus, this factor encodes the curvature of the quantum state space and signifies how much the geodesic path at each instant differes from the actual path that the system travels. 

The quantity $dS(t) -dS_0(t)$ captures the extra evolution of quantum state through the projective Hilbert space.
This may be called quantum circuitousness, because it quantifies the wasted quantum resources when evolution deviates from optimal geodesic paths. This ``winding tax'' can directly impact quantum algorithm efficiency, gate fidelity, and battery charging protocols. By minimizing circuitousness, we can design faster quantum operations that consume less energy while maintaining robustness against noise. It reveals the fundamental trade-off between geometric progress and phase accumulation.

The total distance travelled by the state during the quantum evolution in the interval $[0,T]$ can be expressed as integral over a kernel along the shortest path as given by
\begin{equation}
   S= \int_0^T K(S_0(t), \Phi(t) ) ~~dS_0(t),
\end{equation}
where $K(S_0(t), \Phi(t) )  = \frac{ \sin S_0(t)}{\gamma(t)}  \frac{d\Phi}{dS_0} $ and $\gamma(t) = \sqrt{1- \frac{dS_0^2}{dS^2}} $  is the path contraction factor with $\gamma(t) \le 1$.
The relation suggests that the total distance traveled by the state vector in the Hilbert space is an accumulation of local contributions along the instantaneous geodesic path (the shortest path in the projective space) and these local contributions are modulated by the dynamical and geometric phases.
In other words, the phases cause the state to ``spiral'' or ``twist'' along the geodesic, thereby increasing the total path length. The kernel $K(S_0(t), \Phi(t) )$ quantifies the local rate of this twisting.
Thus, the total distance is the integral of a local factor that depends on the rate of change of the total phase (dynamical and geometric) with respect to the geodesic distance. This factor is always greater than or equal to one, and equals one only when the phase is constant (or changing in a way that the derivative is zero). This reflects the fact that any phase accumulation (whether dynamical or geometric) will cause the state to travel a longer path in the Hilbert space.
Most importantly, this relation suggests that quantum dynamics is completely determined by the minimal distance plus phase information. The ``excess path length'' is encoded entirely in how phases accumulate along the geodesic.
This brings a unification of geometric and dynamical aspects of quantum mechanics, showing that the apparent complexity of quantum evolution reduces to a simple geometric integral.


\section{Quantum Speed Limit and Geometric Phase }

Here, we will prove a new quantum speed limit that depends on the geometric phase acquired by the system.
From Eq.~\eqref{eq:metric}, if we drop the second term, we can have $dS \ge dS_0$, and for a system evolving under a time-independent Hamiltonian, that will lead to the Mandelstam-Tamm \cite{mt}  quantum speed limit bound, i.e., 
$$T \ge \frac{\hbar S_0(T)}{2 \Delta H}, $$
where $S_0(T) = 2 \cos^{-1} |\langle \Psi(0)|\Psi(T) \rangle| $.
Here, we will prove a new quantum speed limit that depends on the time-average of the geometric phase fraction and the geometric phase acquired during the time evolution. First, note that in the controlled-rotating frame, we can express the infinitesimal Fubini-Study metric as
\begin{align}
dS^2(t)  = dS_0(t)^2 + 4 d\Phi_D(t) d\Phi_G(t).
\end{align}
It may be noted that the contributions to the dynamical phase and the geometric phase can depend on the rotating frame \cite{klaus}.
However, we are not exploring this in detail here. We focus on the application of the metric structure in proving quantum speed limit.
Now, if we drop the first term and rearrange, we have 
\begin{align}
\frac{\Delta H }{\hbar} \sqrt \frac{d\Phi_G(t) }{d\Phi_D(t)}  dt   \ge  d\Phi_G(t).
\end{align}
On defining $F = \sqrt \frac{d\Phi_G(t) }{d\Phi_D(t)}$ we have the time average of square root of the geometric phase 
fraction with respect to the dynamical phase as 
\begin{align}
{\bar F} = \frac{1}{T} \int_0^T  \sqrt \frac{d\Phi_G(t) }{d\Phi_D(t)}  dt  = 
\frac{1}{T} \int_0^T  \tan \frac{S_0(t)}{2}  dt .
\end{align}
On integrating both the sides, we have 
\begin{align}
T \ge \frac{\hbar |\Phi_G(T)| }{\Delta H \bar F}.
\end{align}

This can be regarded as the geometric phase induced quantum speed limit, and it is independent of any QSL bounds known to date.
The Mandelstamm-Tam bound on the QSL come from quantum dynamics, where as our bound comes from pure geometric phase.  The geometric phase $\Phi_G(T)$ accumulates as the quantum state traverses an open path in the projective Hilbert space. Unlike the conventional QSL based on the Bures angle (which depends only on the initial and final states), the geometric phase is path-dependent. This means that if we have more winding in state space, that can tighten the QSL. By incorporating the geometric phase, the new QSL provides a precise figure of merit revealing the extent of non-geodesic evolution undertaken  by the quantum state, quantifying how ``circuitous'' its path was. The bound suggests that the minimum time for a quantum process is determined by the ratio of the geometric path complexity to the speed of quantum evolution (apart from other factors). The geometric complexity of paths is fundamentally different from conventional QSLs that depend only on the initial and final states.

Our result not only provides a new bound for navigating trade-offs between quantum speed limit and stability (high geometric fraction), but also provide a new geometric phase dependent quantum speed limit for cyclic evolutions.
The standard Mandelstam-Tamm bound becomes trivial for cyclic evolutions, since the initial and final states are identical (up to a phase), making the Bures angle $S_0(T) = 0$. This renders the bound physically useless.
The quantum speed limit, which incorporates the geometric phase, provides a new direction hitherto unnoticed. For a cyclic evolution where the state returns to its initial ray in the projective Hilbert space, we have $|\langle \Psi(0)|\Psi(T)\rangle| = 1$. In such cases, the accumulated geometric phase $\Phi_G$ becomes the non-trivial measure of the path-dependent factor for the quantum system.
A speed limit of the form 
$T \ge \frac{\hbar |\Phi_G(T)| }{\Delta H \bar F}$
provides a fundamental lower bound on the time required to complete the cycle based on the holonomy, or the ``anholonomy,'' of the path.
This is important for optimizing cyclic processes like geometric quantum gates, where the operation is defined by the geometric phase, and for understanding the quantum dynamics of driven systems. This can also be interpreted as how the system's dynamical resources (energy fluctuations) plays a role in setting a speed limit for generating geometric phases.

\section{Conclusion}

To summarize, we have presented a fundamental result that quantitatively relates the fractional contributions of dynamical and geometric phases to the total phase in terms of the geometric distance from the initial state. This provides a universal law connecting geometry and dynamics in quantum evolution, powerful tool for quantum engineering and control, new insights into the fundamental structure of quantum mechanics and testable predictions for experimental verification.
We have defined the geometric phase fraction (GPF) which can quantitatively answers the question: ``How geometric is the quantum evolution?''
From a practical standpoint, the GPF is a key figure of merit for quantum engineering and serving for the design of robust technologies. In quantum computation, maximizing the GPF during a gate operation becomes a concrete strategy for actively building in resilience to control noise and energy fluctuations, rather than treating it as a passive outcome.  
We have proved a new quantum speed limit that depends on the geometric phase and can be applied to cyclic quantum evolutions.
We believe that our results will openup new avenues for geometric quantum computing, optimal control of quantum dynamics, quantum speed limit and its applications to quantum information processing. 



\section{Appendix}

\section{Proof of Lemma}

We have a unitary operator $U$ and a normalised state $|\Psi \rangle$. The claim is:
\[
U |\Psi \rangle = \langle U \rangle |\Psi \rangle + \Delta U \ |{\bar \Psi}(U) \rangle,
\]
where $\langle U \rangle = \langle \Psi  | U  | \Psi \rangle $,  
$\Delta U = \sqrt{\langle U^\dagger U \rangle - |\langle U \rangle|^2}$ and
$|{\bar \Psi}(U) \rangle$ is some normalized state orthogonal to $|\Psi \rangle$.

\section{Proof}

Any vector $U|\Psi \rangle$ can be decomposed into a component along $|\Psi \rangle$ and a component orthogonal to $|\Psi \rangle$.
Define

\begin{align}
|\Psi_{\parallel}\rangle &= \langle \Psi | U |\Psi \rangle \, |\Psi \rangle, \\
|\Psi_{\perp}\rangle &= U|\Psi \rangle - \langle\Psi | U | \Psi \rangle \, |\Psi \rangle.
\end{align}
Clearly $\langle\Psi |\Psi_{\perp}\rangle = 0$ and $|\Psi_{\perp}\rangle $ is not normalised. Therefore, we have
\[
U|\Psi \rangle = \langle U\rangle |\Psi \rangle + |\Psi_{\perp}\rangle.
\]
 To find the norm of $|\Psi_{\perp}\rangle$, we compute
\[
\||\Psi_{\perp}\rangle\|^2 = \langle\Psi_{\perp}|\Psi_{\perp}\rangle
\]
\[
= \langle\Psi | U^\dagger U |\Psi \rangle - \langle U\rangle^* \langle\Psi |U |\Psi \rangle - \langle U\rangle \langle \Psi |U^\dagger|\Psi \rangle + |\langle U\rangle|^2.
\]
Also $\langle\Psi |U^\dagger|\Psi \rangle = \langle U\rangle^*$. Therefore, we have 
\[
\||\Psi_{\perp}\rangle\|^2 = 1 - |\langle U\rangle|^2.
\]

The uncertainty in $U$ is defined as $\Delta U = \sqrt{ \langle U^\dagger U\rangle - |\langle U\rangle|^2 }$. 
Since $\langle U^\dagger U\rangle = 1$, we have
\[
\Delta U = \sqrt{1 - |\langle U\rangle|^2}.
\]
Using $\||\psi_{\perp}\rangle\| = \sqrt{1 - |\langle U\rangle|^2} = \Delta U$,  we can write
\[
|\Psi_{\perp}\rangle = (\Delta U) \, |{\bar \Psi}(U) \rangle,
\]
where $|{\bar \Psi}(U) \rangle$ is the normalized version of $|\Psi_{\perp}\rangle$, i.e.,
\[
|{\bar \Psi}(U) \rangle  = \frac{|\Psi_{\perp}\rangle}{\Delta U}
\]
provided $\Delta U \neq 0$. If $\Delta U = 0$, then $|\Psi_{\perp}\rangle = 0$ and $|{\bar \Psi}(U)  \rangle$ can be arbitrary (orthogonal to $|\Psi \rangle$). Thus, we have 
\[
U |\Psi \rangle = \langle U\rangle \, |\Psi \rangle + \Delta U \, |{\bar \Psi}(U) \rangle
\]
with $\langle \Psi |{\bar \Psi}(U)\rangle = 0$ and $\langle {\bar \Psi}(U) | {\bar \Psi}(U) \rangle = 1$.
Also, note that $|{\bar \Psi}(U) \rangle$ depends on the unitary $U$.

\section{Proof of Fractional Contributions in Eq(9)}

Consider the state at time t as given by

\begin{align}
|\Psi(t)\rangle = \cos \frac{S_0(t)}{2} e^{i \Phi(t)} | \Psi(0) \rangle + \sin \frac{S_0(t)}{2}
 | {\bar \Psi}_0(t) \rangle ,
 \end{align}
 where $|{\bar \Psi}_0(t) \rangle$ is orthogonal to $| \Psi(0) \rangle$. Since $|{\bar \Psi}_0(t) \rangle$  depends on $U(t)$, the time dependence will come from $U(t)$ dependence.

First consider the derivative of the dynamical phase as given by
\begin{align}
\frac{d\Phi_D }{dt}= - \frac{1}{\hbar} \langle \Psi(t) |H | \Psi(t) \rangle = -i \langle \Psi(t) |{\dot \Psi(t)} \rangle.
\end{align}

Now, let us calculate $\frac{d\Phi_D }{dt}$.
\begin{eqnarray}
\frac{d\Phi_D }{dt} & = -i \langle \Psi(t) |{\dot \Psi(t)} \rangle = \cos^2 \frac{S_0(t)}{2}  \frac{d \Phi(t)}{dt} \\ \nonumber
& -i ~\cos \frac{S_0(t)}{2} \sin \frac{S_0(t)}{2} e^{-i\Phi(t) }
\langle \Psi(0) | {\dot {\bar \Psi}_0(t)} \rangle \\ \nonumber
&- i ~ \sin^2 \frac{S_0(t)}{2}
\langle {\bar \Psi_0(t)} | {\dot {\bar \Psi}_0(t)} \rangle.
\label{eq:dD}
\end{eqnarray} 
From the conditions $\langle \Psi(0) | {\bar \Psi}_0(t) \rangle =0$, we have $\langle \Psi(0) | {\dot {\bar \Psi}_0(t)} \rangle = 0 $. 
Then, Eq.~\eqref{eq:dD} can be expressed as 
\begin{align}
\frac{d\Phi_D }{dt}  = \cos^2 \frac{S_0(t)}{2}  \frac{d \Phi(t)}{dt} - i ~ \sin^2 \frac{S_0(t)}{2}
\langle {\bar \Psi_0(t)} | {\dot {\bar \Psi}_0(t)} \rangle.
\end{align} 

Using the normalization condition, i.e.,  $\langle {\bar \Psi_0(t)} | {\bar \Psi}_0(t) \rangle =1 $, it follows that  
$\langle \bar{\Psi}_0(t) | \dot{\bar{\Psi}}_0(t) \rangle$ is purely imaginary. Hence, $-i \langle {\bar \Psi_0(t)} | {\dot {\bar \Psi}_0(t)} \rangle$ is real. Let us denote the phase acquired by the orthogonal component as
\begin{eqnarray}
{\bar \Phi} = -i \int \langle {\bar \Psi_0(t)} | {\dot {\bar \Psi}_0(t)} \rangle dt.   
\end{eqnarray}

Now the rate of change of the dynamical phase can be expressed as 

\begin{align}
\frac{d\Phi_D }{dt}  = \cos^2 \frac{S_0(t)}{2}  \frac{d \Phi(t)}{dt} 
+ \sin^2 \frac{S_0(t)}{2}  \frac{d {\bar \Phi}}{dt}.
\end{align} 

Therefore, we can write the infinitesimal change in the dynamical phase as 
\begin{align}
d\Phi_D  = \cos^2 \frac{S_0(t)}{2}  d \Phi(t) 
+ \sin^2 \frac{S_0(t)}{2}  d {\bar \Phi}.
\end{align} 

Similarly, the expression for the infinitesimal change in the geometric phase is given by 
\begin{align}
d\Phi_G  = \sin^2 \frac{S_0(t)}{2}  d \Phi(t) 
- \sin^2 \frac{S_0(t)}{2}  d {\bar \Phi}.
\end{align}

We can interpret this as the fractional contributions of total phase and phase coming from orthogonal part to the dynamical and the geometric phases. But the fraction is still controlled by single geometric quantity, i.e., the Bures angle $S_0(t)$.

\end{document}